\documentclass[manuscript,screen]{acmart}

\usepackage{pifont}
\usepackage{enumitem}
\usepackage{multirow}
\usepackage{colortbl}
\usepackage{framed}
\usepackage{booktabs}
\usepackage{adjustbox}
\usepackage[most]{tcolorbox}
\usepackage{listings}
\usepackage{xcolor}

\definecolor{lightGray}{gray}{0.92}
\definecolor{deepGray}{RGB}{64,64,64}
\definecolor{lightGreen}{RGB}{144, 238, 144}
\definecolor{lightBlue}{RGB}{235, 245, 251} 
\definecolor{deepGreen}{RGB}{0,128,0}
\definecolor{deepRed}{RGB}{178,34,34}
\definecolor{shadecolor}{RGB}{245,245,245}

\newenvironment{custommdframed}
{%
    \MakeFramed{\advance\hsize-6pt\FrameRestore}%
    \noindent%
}
{\endMakeFramed}
\AtBeginDocument{%
  }

\newcommand{\ourmethod}{\textsc{ReConMem}}

\begin{document}

\title{\ourmethod{}: A Graph-Based Dual Memory System for Repository-Oriented Conversations}

\author{Zhishuo Yang}
\email{yangzhishuo@buaa.edu.cn}
% \orcid{0009-0006-9836-409X}
\affiliation{%
  \institution{State Key Laboratory of Complex \& Critical Software Environment, School of Computer Science and Engineering, Beihang University}
  \city{Beijing}
  \country{China}
}

\author{Li Zhang}
\email{lily@buaa.edu.cn}
\affiliation{%
  \institution{State Key Laboratory of Complex \& Critical Software Environment, School of Computer Science and Engineering, Beihang University}
  \city{Beijing}
  \country{China}
}

\author{Yang Liu}
\email{liuyang26@buaa.edu.cn}
\affiliation{%
  \institution{State Key Laboratory of Complex \& Critical Software Environment, School of Computer Science and Engineering, Beihang University}
  \city{Beijing}
  \country{China}
}

\author{Donghao Yang}
\email{yangdonghao@buaa.edu.cn}
\affiliation{%
  \institution{State Key Laboratory of Complex \& Critical Software Environment, School of Computer Science and Engineering, Beihang University}
  \city{Beijing}
  \country{China}
}

\author{Fang Liu}
\email{fangliu@buaa.edu.cn}
\correspondingauthor
\affiliation{%
  \institution{State Key Laboratory of Complex \& Critical Software Environment, School of Computer Science and Engineering, Beihang University}
  \city{Beijing}
  \country{China}
}

\renewcommand{\shortauthors}{Yang et al.}


\begin{abstract}
Large language model (LLM)-based code assistants are increasingly used in iterative repository-level development, where effective assistance requires jointly leveraging evolving conversational history and repository context. Existing memory systems, however, are primarily designed for general-purpose conversations and do not sufficiently coordinate conversational information with the structured code entities to which it refers. We propose \ourmethod{}, a graph-based dual memory system for repository-oriented conversations. 
\ourmethod{} organizes distilled conversational memories and conversation-relevant code entities into interconnected semantic and structural memory graphs, while maintaining their cross-graph associations as interactions evolve. For memory recall, \ourmethod{} combines natural-language and code-aligned retrieval through code-draft-based query grounding, expands retrieved information over intra- and inter-graph relations, and compresses retrieved code according to its structural granularity. 
We evaluate \ourmethod{} on two repository-oriented conversational benchmarks covering controlled long-horizon context management and realistic project-grounded multi-turn assistance. On LoCoEval, \ourmethod{} improves overall Pass@1 score over the strongest baseline by 7.917 and 7.292 percentage points on the single-hop and multi-hop subsets, respectively, while maintaining comparable context compression. On CodeAssistBench, it achieves 27.61\% Correct and 44.03\% Partially Correct, outperforming Full History by 4.10 and 7.09 percentage points, respectively, while also improving interaction quality. Further analyses demonstrate the complementary contributions of the two retrieval routes and the effectiveness of \ourmethod{} across different conversation characteristics. These results show that explicitly coordinating conversational history with repository structure enables more effective context management for repository-oriented multi-turn interactions.
We provide the replication package at \url{https://github.com/CoconutTreeee/ReConMem}.
\end{abstract}





\maketitle

\section{Introduction}
Modern software development increasingly involves interactive collaboration with AI code assistants \cite{Cursor, Trae, Cline, ClaudeCode, GitHub-Copilot} driven by large language models (LLMs).
Beyond single-turn code tasks, developers now engage in iterative interactions with these assistants in repository-level development, such as code understanding, feature implementation, and refactoring, resulting in overwhelming context that exceeds the LLM's capacity.
This raises a central challenge: how to effectively maintain, associate, and utilize context.

Recent advances in memory systems have been extensively explored to manage conversational context and maintain stable interactions across turns \cite{packer2023memgpt, chhikara2025mem0, zhong2024memorybank, tan2025prospect}.
Rather than loading the entire context into the LLM prompt, these systems maintain external storage and retrieve only task-relevant information on demand.
For example, Mem0 dynamically extracts salient information from conversations and employs graph-based memory representations to capture complex relational structures, thereby improving retrieval.

\begin{figure}[t]
    \centering
    \setlength{\abovecaptionskip}{0.1cm}
    \includegraphics[width=0.65\linewidth]{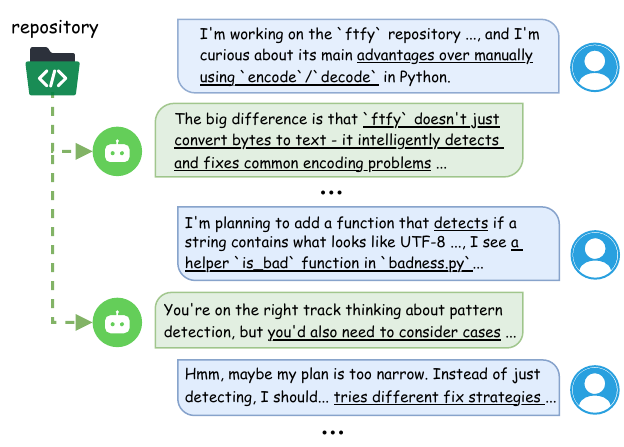}
    \caption{A typical interaction between a user and a code assistant, involving cross-file reasoning, iterative requirements, and long conversational history.}
    \label{fig:intro_conv_exp}
\end{figure}

However, these methods are designed for general scenarios and fail to fully exploit the characteristics of repository-level development.
Conversation and repository are inherently interdependent here: requirements, decisions, and clarifications discussed across turns are often tied to specific repository entities, while the repository context relevant to a new request may in turn depend on the preceding interaction history.
As illustrated in Figure \ref{fig:intro_conv_exp}, the developer begins with a question about \texttt{ftfy} versus manual encode/decode, proposes a detection function based on \texttt{is\_bad} in \texttt{badness.py}, and then revises the plan toward different fix strategies after the assistant identifies missing cases.
The later request builds on both the earlier discussion and the shared repository entity \texttt{is\_bad} in \texttt{badness.py}, showing how conversational history and repository context are intertwined.
Since existing methods fail to sufficiently coordinate these two types of context when organizing, retrieving, and utilizing context, their performance remains unsatisfactory.

Effectively handling such intertwined context introduces several concrete challenges.
Conversational memories and repository entities need to be associated rather than maintained as isolated context; retrieval needs to jointly leverage conversational semantics and information encoded in repository code, despite their heterogeneous representations; 
and retrieved code needs to retain sufficient task-relevant information without introducing excessive irrelevant details, even when relevant code entities vary substantially in granularity.

To address these challenges, we propose \ourmethod{}, a graph-based dual memory system that represents and utilizes two sources of context through two aligned yet distinct memory spaces. 
Specifically, \ourmethod{} maintains a \textbf{Semantic Memory Graph} that captures distilled natural language memories and their relations from conversations, and a \textbf{Structural Memory Graph} that maintains conversation-relevant code entities and organizes them according to the repository structure.
These two memory spaces are explicitly linked through dual relations and kept consistent by a dual synchronization protocol. Together, they form a unified yet dual-view representation, which we refer to as the dual memory graph. 
Building on this representation, we introduce a joint memory recall mechanism. Considering the heterogeneity of the two graphs, \ourmethod{} employs a code-draft-based query grounding strategy to derive a code-aligned representation of the original query. The original natural-language query and the resulting dual query are then used to retrieve from the semantic and structural memory graphs, respectively. 
We further apply graph-adjacency expansion over intra- and inter-graph relations to retrieve additional relevant context.
For the retrieved code, we propose a hierarchical code context compression method that applies granularity-specific compression strategies to reduce task-irrelevant information while preserving task-relevant content.
%%%%%%%%%%%%%%%%%%%%%%%%%%%%%%%%%%%%%%
We evaluate \ourmethod{} on two repository-oriented conversational benchmarks, LoCoEval~\cite{liu2026scalable} and CodeAssistBench (CAB)~\cite{kim2025codeassistbench}, covering both controlled long-horizon context management and realistic project-grounded multi-turn assistance.
On LoCoEval, \ourmethod{} improves the overall Pass@1 over the strongest baseline by 7.917 and 7.292 percentage points on the single-hop and multi-hop subsets, respectively, while maintaining a comparable context compression ratio.
On CAB, \ourmethod{} achieves 27.61\% Correct and 44.03\% Partially Correct, outperforming Full History by 4.10 and 7.09 percentage points, respectively; it also achieves the highest Appropriate rate and the fewest average interaction turns among the compared methods.
Further analyses verify the effectiveness of individual components, reveal the complementary roles of the two retrieval branches, and demonstrate the efficiency and robustness of \ourmethod{} under different repository-oriented conversational settings.

In summary, our main contributions are as follows:
\begin{itemize}
    \item We propose \ourmethod{}, a dual-graph memory system specifically designed for repository-oriented conversations. It introduces a dual-view memory organization that jointly represents conversational history and repository code, maintaining complementary context from two perspectives.
    \item We design a joint memory recall mechanism tailored to the heterogeneous representations of conversational and code contexts. 
    By combining code-draft-based query grounding and graph-adjacency-expanded retrieval, \ourmethod{} retrieves comprehensive and diverse contextual information.
    \item We conduct systematic evaluations of \ourmethod{} on LoCoEval and CodeAssistBench, covering both controlled long-horizon context management and realistic project-grounded multi-turn programming assistance. 
    The results demonstrate the effectiveness of \ourmethod{} by achieving consistent improvements over strong baselines.
\end{itemize}

\section{Methodology}
\begin{figure*}[t]
    \centering
    \setlength{\abovecaptionskip}{0.1cm}
    \includegraphics[width=0.8\linewidth]{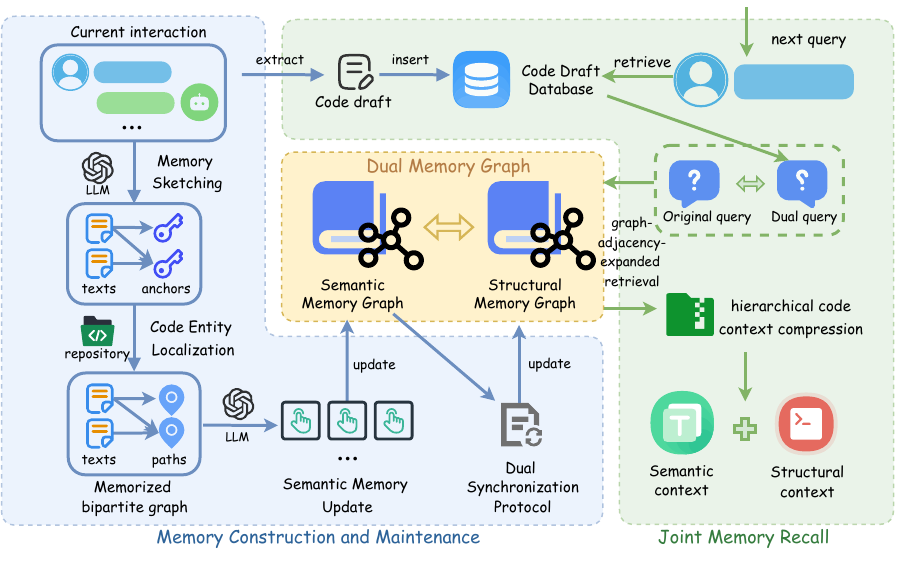}
    \caption{Overview of \ourmethod{}.}
    \label{fig:overview}
    \vspace{-0.3cm}
\end{figure*}

As shown in Figure \ref{fig:overview}, the core memory structure of \ourmethod{} is the \textbf{Dual Memory Graph},
which consists of a semantic memory graph and a structural memory graph, connected to each other through dual relations.
The workflow of the entire method is centered around the life cycle of the dual memory graph: after each interaction, it enters the \textbf{Memory Construction and Maintenance} stage, where it identifies and structures information worth retaining from the current turn, 
and consolidates these interaction-level memory sketches into the persistent dual memory graph;
when a new user query arrives, it enters the \textbf{Joint Memory Recall} stage, where it retrieves and processes complementary conversational and code context from the maintained memory structures to support response generation.

\subsection{Dual Memory Graph}
\label{sec:dual_memory_graphs}

\begin{figure}[t]
    \centering
    \setlength{\abovecaptionskip}{0.2cm}
    \includegraphics[width=0.7\linewidth]{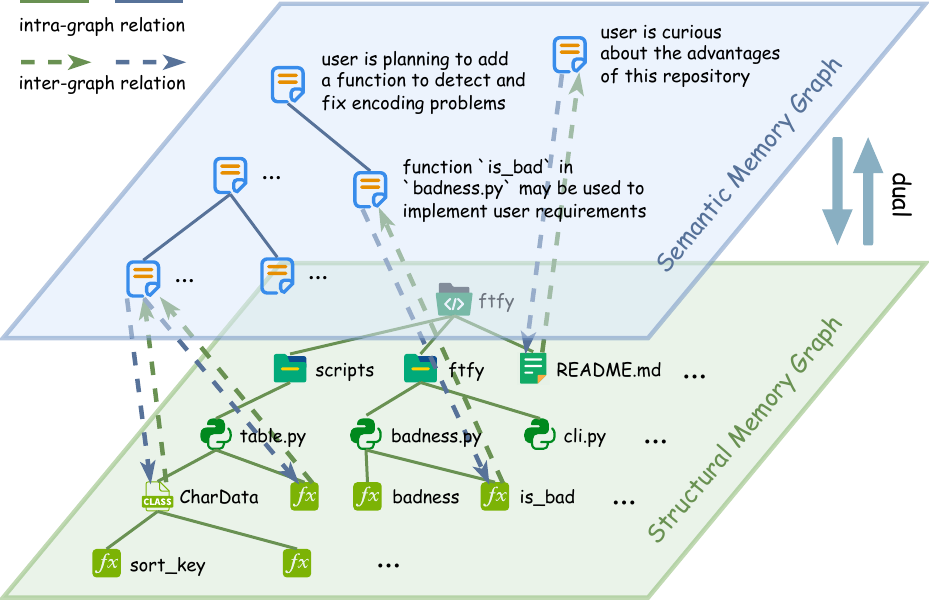}
    \caption{Illustration of the dual memory graph structure.}
    \label{fig:dual_memory_graphs}
    \vspace{-0.3cm}
\end{figure}

As shown in Figure \ref{fig:dual_memory_graphs}, the dual memory graph organizes conversational memories and repository code entities in two distinct but interconnected memory spaces: a semantic memory graph and a structural memory graph. Formally, we define it as
\begin{equation}
G_{\mathrm{dual}}
=
(G_{\mathrm{se}}, G_{\mathrm{st}}, R_{\mathrm{dual}}, R_{\mathrm{dual}}^{*}),
\end{equation}
where $G_{\mathrm{se}}=(E_{\mathrm{se}},R_{\mathrm{se}})$ and $G_{\mathrm{st}}=(E_{\mathrm{st}},R_{\mathrm{st}})$ denote the semantic and structural memory graphs, respectively. $E_{\mathrm{se}}$ and $E_{\mathrm{st}}$ are their entity sets, while $R_{\mathrm{se}}$ and $R_{\mathrm{st}}$ represent the corresponding intra-graph relations. $R_{\mathrm{dual}}\subseteq E_{\mathrm{se}}\times E_{\mathrm{st}}$ and $R_{\mathrm{dual}}^{*}\subseteq E_{\mathrm{st}}\times E_{\mathrm{se}}$ represent the inter-graph relations connecting the two memory spaces.

In the semantic memory graph, each entity $e_{\mathrm{se}}\in E_{\mathrm{se}}$ corresponds to a textual memory %derived from the textual descriptions in the memorized bipartite graphs. 
extracted from preceding interactions and maintained through the construction process described in \S \ref{sec:memory_construction_and_maintenance}.
The directed intra-graph relations $R_{\mathrm{se}}$ are dynamically established during the memory updating process (\S \ref{sec:semantic_memory_update}) to capture meaningful associations between different memory entities. 
These relations preserve local associations among otherwise separate textual memories and provide the semantic neighborhood used in subsequent graph-based retrieval. 

In the structural memory graph, each entity $e_{\mathrm{st}}\in E_{\mathrm{st}}$ corresponds to a repository code entity currently associated with at least one textual memory.  
The structural memory graph is constructed selectively, containing only code entities referenced by the semantic memories, including files, type declarations, functions or methods, and variables, fields, or attributes.
Each structural entity retains its complete repository path, from which its structural relation can be derived. Let $\pi(e)$ denote the repository path of a structural entity $e$. The directed structural relation set is defined as
\begin{equation}
R_{\mathrm{st}}
=
\left\{
(e_i,e_j)\in E_{\mathrm{st}}\times E_{\mathrm{st}}
\mid
\operatorname{parent}(\pi(e_i))=\pi(e_j)
\right\}.
\end{equation}
Accordingly, $R_{\mathrm{st}}$ captures direct child-to-parent containment relations only between repository entities that have both been instantiated in the structural memory graph, forming a sparse structural view of the repository conditioned on the currently maintained memories.

The two memory graphs are connected through dual relations that explicitly associate textual memories with the repository entities to which they refer. For a semantic entity $e_{\mathrm{se}}$ with one or more valid code associations, $R_{\mathrm{dual}}$ contains directed relations from $e_{\mathrm{se}}$ to the corresponding structural entities, while $R_{\mathrm{dual}}^{*}$ maintains their reverse mappings. A semantic entity may therefore be associated with multiple structural entities, and the same structural entity may be referenced by multiple semantic entities. Textual memories without valid code associations remain solely in the semantic memory graph. The two relation sets satisfy
\begin{equation}
R_{\mathrm{dual}}
=
\left\{
(e_{\mathrm{se}},e_{\mathrm{st}})
\mid
(e_{\mathrm{st}},e_{\mathrm{se}})
\in R_{\mathrm{dual}}^{*}
\right\},
\label{eq:dual_relation}
\end{equation}
where $e_{\mathrm{se}}\in E_{\mathrm{se}}$ and $e_{\mathrm{st}}\in E_{\mathrm{st}}$. This bidirectional mapping preserves the associations between conversational memories and their referenced repository entities, allowing subsequent retrieval to traverse between the semantic and structural memory spaces.

\subsection{Memory Construction and Maintenance}
\label{sec:memory_construction_and_maintenance}
In this phase, we extract key information from the current turn and represent it as textual descriptions together with their associated code-entity anchors. We then perform code entity localization to resolve these anchors into exact entity paths, thereby forming a series of \textbf{Memorized Bipartite Graph}s. Next, to integrate these dispersed memorized bipartite graphs into a unified memory structure, namely the dual memory graph, we introduce an LLM-driven decision mechanism to perform \textbf{Semantic Memory Update}. Subsequently, we employ a \textbf{Dual Synchronization Protocol} to consistently propagate the corresponding updates to the Structural Memory Graph, thereby maintaining the bidirectional consistency of the cross-graph mappings.

\subsubsection{Memorized Bipartite Graph}
\label{sec:memorized_bipartite_graph}

For each interaction turn, we use an LLM to construct a bipartite graph to organize textual memory units and their associated code entities. 
Formally, for each interaction turn $t$, we construct
\begin{equation}
    \mathcal{B}_t = (D_t, P_t, R_t),
\end{equation}
where $D_t = \{d_i\}_{i=1}^{M}$ and $P_t = \{p_j\}_{j=1}^{N}$ are two disjoint vertex sets.
$D_t$ contains textual memory units extracted from the interaction, which serve as candidates for subsequent updates into the Semantic Memory Graph, while $P_t$ contains repository-grounded code entities involved in these memories, which correspond to units in the Structural Memory Graph.
The directed edge set $R_t \subseteq D_t \times P_t$ records which code entities are involved in each textual memory.
In this way, the bipartite graph serves as a turn-level bridge between the current interaction and the two memory spaces, providing the basis for subsequent memory integration and cross-graph association.

The construction of the bipartite graph is guided by two considerations tailored to repository-oriented development: retaining information that may remain useful in subsequent development activities, and keeping each memory unit sufficiently fine-grained for independent retrieval and update.
For the textual vertex set $D_t$, we prompt an LLM to extract development-relevant information from the current user--assistant turn.
The extracted information broadly covers requirement-oriented information, such as target functionality, structural constraints, dependencies, coding conventions, and input/output expectations, as well as contextual information relevant to implementation, such as project context, design decisions, task intentions, and code-specific observations.
Each textual vertex $d_i \in D_t$ is kept atomic and concise, typically representing a single development intent, requirement, constraint, decision, or code-related observation.

Along with each textual memory $d_i$, the LLM also produces a set of candidate code-entity anchors, denoted by $\tilde{A}(d_i)$, representing repository-style references associated with that memory.
Explicitly mentioned repository entities, such as files, modules, classes, and functions, are prioritized when constructing these anchors.
When precise code locations are unavailable, a small number of high-confidence approximate references may also be inferred from the entity mentions and the local interaction context.
Consequently, the resulting anchors may be incomplete or approximate hierarchical references, such as \texttt{CharData/sort\_key}, rather than exact repository locations.

Because these candidate anchors are derived from conversational information, they may not exactly correspond to valid entities in the actual repository structure. We therefore perform code entity localization to resolve each candidate anchor $\tilde{a} \in \tilde{A}(d_i)$ into at most one concrete code-entity path in the repository. We denote this process by $L(\tilde{a})$, where $L(\tilde{a})=\bot$ if no valid entity can be resolved.

Specifically, we adopt a two-stage resolution strategy. We first perform an exact lookup over the repository structure. If the anchor cannot be resolved exactly, we collect non-directory code entities whose terminal path component exactly matches that of the anchor and rank them according to their structural agreement with the anchor path. The ranking prioritizes candidates with longer consecutively matched trailing path components, followed by those with longer consecutively matched leading path components; the string length of the candidate path is used only as a final tie-breaker. The highest-ranked candidate is selected as the localized path, while anchors with no valid candidate remain unresolved and are discarded.

For each textual memory $d_i$, the successfully resolved entities constitute
\begin{equation}
A(d_i)=
\left\{
L(\tilde{a})
\mid
\tilde{a}\in\tilde{A}(d_i),
L(\tilde{a})\neq\bot
\right\}.
\label{eq:localized_entities}
\end{equation}

Accordingly, the code-entity vertex set and the cross-set edge set of the memorized bipartite graph are defined as
\begin{equation}
P_t=
\bigcup_{d_i\in D_t}A(d_i),
\label{eq:bipartite_code_vertices}
\end{equation}
and
\begin{equation}
R_t=
\left\{
(d_i,p_j)\in D_t\times P_t
\mid
p_j\in A(d_i)
\right\}.
\label{eq:bipartite_relations}
\end{equation}

A textual vertex is not required to be associated with a code entity, i.e., $A(d_i)$ may be empty, whereas every code-side vertex in $P_t$ is associated with at least one textual vertex. 

\subsubsection{Semantic Memory Update}
\label{sec:semantic_memory_update}

To consolidate overlapping information and avoid redundant or fragmented memory accumulation, we employ an LLM-driven mechanism to update the semantic memory graph based on the memorized bipartite graph.

Specifically, for each textual description $d_i$ in the memorized bipartite graph, we first retrieve the top-$k_1$ most relevant textual entities from the current semantic memory graph according to embedding similarity, together with the paths of their connected code entities. 
These retrieved entities and code paths serve as reference context for the update decision. We then feed this context together with the memorized bipartite graph into an LLM and formulate semantic memory update as a structured decision task. The model jointly determines an appropriate update action from a predefined set $\{\textsc{Insert}, \textsc{Modify}, \textsc{Remove}\}$ and the semantic adjacency relations associated with the update.

Semantic adjacency relations are established between memories concerning different but related development entities or topics. In contrast, information concerning the same referent is consolidated through memory updates rather than represented as an adjacency relation. Based on these principles, the three update actions are defined as follows:
\begin{itemize}
    \item \textbf{INSERT}: If $d_i$ is semantically unrelated to existing textual memory entities or corresponds to a different referent, this action is performed, instantiating $d_i$ as a new entity in the semantic memory graph. 
    Its textual content and code associations are initialized from the corresponding information in the memorized bipartite graph, while directed semantic adjacency relations may be established with related existing memories.
    \item \textbf{MODIFY}: If $d_i$ is semantically similar to an existing textual memory entity and corresponds to the same referent, but provides supplementary or updated information, this action is applied. 
    The textual content is consolidated with the new information, while the existing and newly introduced code associations are merged. Its semantic adjacency relations are updated accordingly.
    \item \textbf{REMOVE}: If $d_i$ exhibits a clear semantic conflict with an existing textual memory entity, or if the user explicitly invalidates prior information, this action is performed to remove the corresponding entity.
\end{itemize}

Through these updates, outdated information is promptly revised or removed, while related information is consolidated or connected through semantic relations, keeping the semantic memory graph non-conflicting and well organized.

\subsubsection{Dual Synchronization Protocol}
\label{sec:dual_synchronization_protocal}
Since the updates described above are directly applied only to the semantic memory graph, we introduce a dual synchronization protocol to consistently propagate these updates to the structural memory graph.
Specifically, when an update introduces a new association between a semantic entity $e_{\mathrm{se}}$ and a repository code entity $e_{\mathrm{st}}$, the corresponding structural entity is instantiated if it does not already exist, and the reciprocal cross-graph mappings are established. Conversely, when an association is removed, the corresponding mappings are removed; if a structural entity is no longer referenced by any semantic entity, it is removed from the structural memory graph.

Through this process, the structural memory graph contains only repository entities referenced by the currently maintained semantic memories, while the cross-graph mappings preserve the bidirectional consistency defined in Eq.~\ref{eq:dual_relation}. This allows subsequent retrieval to traverse consistently between the semantic and structural memory spaces.

\subsection{Joint Memory Recall}
\label{sec:joint_memory_recall}

In this phase, for each user query, we first perform \textbf{Code-Draft-Based Query Grounding} to obtain a code-modality representation aligned with the structural memory graph. This representation serves as the dual query and, together with the original query, is applied to the dual memory graph for similarity-based retrieval over the structural and semantic memory graphs, respectively, yielding a set of initially retrieved entities. Based on these initial results, we adopt a \textbf{Graph-Adjacency-Expanded Retrieval} strategy to expand the retrieval set through intra-graph and inter-graph adjacency relations. The retrieved entities are then merged and deduplicated, after which the retrieved code entities are processed through \textbf{Hierarchical Code Context Compression} to streamline the retrieved context and reduce noisy information. Finally, we obtain memory texts from the textual entities and compressed code from the code entities, forming the semantic and structural contexts, respectively. Together, they provide comprehensive and accurate background information for the code assistant to generate responses to the user.
\subsubsection{Code-Draft-Based Query Grounding}
\label{sec:Code-Draft-Based_Query_Grounding}

In iterative repository-oriented interactions, later user requests often build upon code discussed or generated in preceding turns. 
Consequently, the current user query may not fully express the intended development state in isolation, as part of its meaning needs to be interpreted in conjunction with the preceding interaction context. Meanwhile, code artifacts generated in preceding turns provide a concrete representation of the evolving implementation state. We therefore leverage these code artifacts to ground the current query in the development context established through preceding interactions.

To support this process, we extract code blocks from the responses generated by the code assistant and incrementally store them in an auxiliary code draft database. This database accumulates intermediate code artifacts produced throughout the conversation, serving as a complementary representation of the evolving conversational context in the code modality.

Given a new user query, we retrieve the most similar code snippet from the code draft database based on embedding similarity with the query. The retrieved code snippet provides a code-side representation that is relevant to the current request and grounded in the preceding interaction history. If the highest similarity exceeds a predefined threshold, the retrieved code snippet is used as the dual query for retrieval over the structural memory graph. Otherwise, we fall back to the original user query to avoid introducing an unreliable code-side representation.

\subsubsection{Graph-Adjacency-Expanded Retrieval}
\label{graph_adjacency_expaned_retrieval}

Building on the original user query and the dual query, we first identify a set of initial entities that are directly relevant to the current request from the semantic memory space and the code space through embedding-similarity-based retrieval. However, similarity-based retrieval primarily relies on the representational similarity between the query and individual entities, and may therefore overlook contextual information that is explicitly related to the retrieved entities but does not exhibit sufficiently high similarity to the query itself. To further exploit the dual memory graph structure, we introduce a graph-adjacency-expanded retrieval mechanism that supplements the initial retrieval results through intra-graph and inter-graph relations.

Specifically, on the semantic side, we retrieve the top-$k_2$ most relevant textual entities from the semantic memory graph based on embedding similarity with the original query. On the code side, we divide the retrieval budget $k_3$ into $k_{31}$ and $k_{32}$, i.e., $k_3 = k_{31} + k_{32}$. Among them, $k_{31}$ code entities are retrieved from all code entities in the repository, while the other $k_{32}$ entities are retrieved from the structural memory graph. Both are retrieved based on embedding similarity with the dual query, or with the original query when no code draft satisfies the grounding threshold in code-draft-based query grounding. Repository-wide retrieval enables the system to discover code entities that have not yet been instantiated in the structural memory graph but are directly relevant to the current request, whereas retrieval over the structural memory graph prioritizes repository entities that have already been associated with conversational memories in preceding interactions. These retrieved entities collectively form the initial retrieval set $E_0$.

Starting from $E_0$, we further perform a two-stage adjacency expansion over the dual memory graph. The first stage performs intra-graph expansion to recover local context within each memory space that is directly related to the initially retrieved entities. In the semantic memory graph, expansion along the directed semantic relations $R_{\mathrm{se}}$ supplements other memory information that is different from but related to the retrieved textual memories. Such information may not be surfaced by the initial retrieval due to insufficient direct semantic similarity with the current query, but can be further recalled through the established memory associations. In the structural memory graph, we expand along the structural relations $R_{\mathrm{st}}$. Since $R_{\mathrm{st}}$ represents direct child-to-parent containment relations between instantiated code entities, this expansion supplements fine-grained code entities with their direct higher-level structural context. Through this intra-graph expansion, we obtain $E_1$.

Formally,
\begin{equation}
E_1 = E_0 \cup
\left\{
e \mid
\exists e' \in E_0,\,
(e', e) \in R_{\mathrm{se}} \cup R_{\mathrm{st}}
\right\}.
\label{eq:intra_graph_expansion}
\end{equation}

The second stage performs inter-graph expansion to complement the retrieved context across the semantic and structural memory spaces. Starting from $E_1$, we expand along the dual relations $R_{\mathrm{dual}}$ and $R_{\mathrm{dual}}^{*}$. For a semantic entity, this expansion further recalls its associated repository code entities; for a structural entity, it further recalls the conversational memories that reference that code entity. Because this stage is applied after intra-graph expansion, not only the initially retrieved entities but also the local neighbors discovered in the first stage can introduce corresponding cross-view context through the dual relations. In this way, conversational memories and repository code information complement each other during retrieval rather than being used as two independent sets of retrieval results.

Formally,
\begin{equation}
E_2 = E_1 \cup
\left\{
e \mid
\exists e' \in E_1,\,
(e', e) \in R_{\mathrm{dual}} \cup R_{\mathrm{dual}}^{*}
\right\}.
\label{eq:inter_graph_expansion}
\end{equation}

The expansion process is deliberately bounded to one intra-graph expansion followed by one inter-graph expansion, rather than recursively traversing the dual memory graph. Therefore, only entities that have explicit relations with the initially retrieved entities or their direct local neighbors are further incorporated into the context, allowing ReConMem to exploit graph structure for complementary retrieval while avoiding unconstrained context expansion. For code entities retrieved through $k_{31}$ from the entire repository but not instantiated in the structural memory graph, they are retained as directly retrieved code context but do not participate in subsequent adjacency expansion because no corresponding graph relations are available. Through this two-stage process, the first stage enriches local context within each memory space, while the second stage further complements conversational information with repository code information, resulting in more complete task-relevant context.

\subsubsection{Hierarchical Code Context Compression}
\label{hierachical_code_context_compression}

The results of graph-adjacency-expanded retrieval provide comprehensive task-relevant context for the code assistant to respond to the user. This context often contains a large number of code entities at varying granularities. 
The full content of some code entities is relatively lengthy, and directly incorporating them into the context can introduce substantial noise and overhead, making it necessary to compress the code context.
Considering that in practical development, higher-level code artifacts are primarily concerned with their overall structure rather than detailed logic, we introduce a hierarchical code context compression mechanism. This mechanism leverages the hierarchical structure of the repository and applies differentiated compression strategies to code entities of varying granularities, balancing information retention with context length.

\begin{figure}[t]
    \centering
    \setlength{\abovecaptionskip}{0.2cm}
    \includegraphics[width=\linewidth]{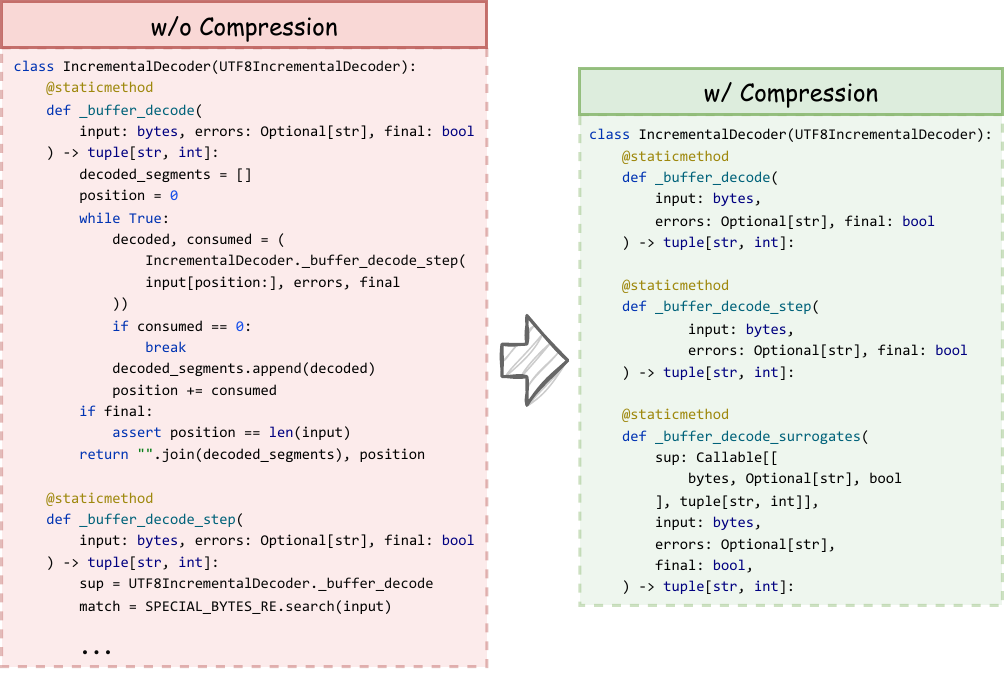}
    \caption{Comparison of a class before and after compression.}
    \label{fig:compression_compare}
    \vspace{-0.3cm}
\end{figure}

Specifically, a function entity, when directly retrieved as a fine-grained execution unit, is typically directly relevant to user requirements and is therefore fully  preserved. In contrast, class and file entities primarily provide structural context, and are thus represented in a compressed form rather than retaining full implementations.
During compression, we perform recursive traversal over the contained entity structure and abstract classes and files into signature-level structured summaries. As shown in Figure \ref{fig:compression_compare}, these summaries preserve key structural information (\textit{e.g.}, interfaces and hierarchical relationships) while omitting implementation details, thereby effectively reducing context length and precisely minimizing redundant noise.

\section{Evaluation}

\subsection{Research Questions}

\begin{itemize}
    \item \textbf{RQ1: Effectiveness.} How effective is \ourmethod{} in managing repository-oriented multi-turn conversations?
    \item \textbf{RQ2: Efficiency.} What is the context-management cost of \ourmethod{}?
    \item \textbf{RQ3: Ablation Study.} To what extent does each key component contribute to \ourmethod{}?
    \item \textbf{RQ4: Context Retrieval Quality.} How effectively does \ourmethod{} retrieve relevant and complementary context?
\end{itemize}

\subsection{Benchmarks}
\label{sec:benchmarks}

To comprehensively evaluate repository-oriented conversational context management,
we select two complementary benchmarks.
LoCoEval~\cite{liu2026scalable} evaluates whether \ourmethod{} can retain and utilize
long-horizon conversational context for repository-level code generation.
CodeAssistBench (CAB)~\cite{kim2025codeassistbench} evaluates whether \ourmethod{} can support realistic interactive
software development assistance over complete repositories.

\subsubsection{LoCoEval}
LoCoEval is constructed from DevEval~\cite{li2024deveval}, a benchmark of manually annotated repository-level code generation tasks.
LoCoEval extracts ground-truth information items from the reference implementations of selected DevEval tasks, introduces distracting information through mutation, and distributes these items into query outlines. During evaluation, a mock-user LLM dynamically generates queries based on the outline and the preceding response, and the evaluated LLM responds using the provided context. After the conversation-generation phase, function-generation queries are issued to evaluate whether the LLM can integrate conversational and repository information to produce implementations that pass the corresponding tests.
LoCoEval contains 320 Python function-generation tasks across 128 long conversations, and is divided into two subsets: single-hop and multi-hop. In the single-hop subset, information relevant to a target task is relatively concentrated within the conversation, whereas in the multi-hop subset, relevant information is distributed across multiple topics and turns, requiring the system to identify and integrate dispersed contextual evidence. Each conversation contains at least 30 but fewer than 70 interaction turns, with an average of approximately 50 rounds. These characteristics make LoCoEval well suited for evaluating long-horizon conversational context management.
The mock-user LLM is \textit{Gemini-2.5-Flash} \cite{comanici2025gemini}.

\subsubsection{CodeAssistBench}
CodeAssistBench (CAB) is a benchmark for multi-turn, project-grounded programming assistance constructed from real-world GitHub issues. 
Unlike LoCoEval, which focuses on Python, the CAB \textit{Verified v3} release used in our experiments contains 274 human-verified issues across seven programming languages---Python, JavaScript, TypeScript, Java, Go, C, and C++---with human-reviewed satisfaction conditions.\footnote{\url{https://github.com/amazon-science/CodeAssistBench}}
CAB's evaluation framework consists of three agents: a User Agent, a Maintainer Agent, and a Judge Agent. 
The User Agent initiates the conversation with a programming question from a GitHub issue and provides follow-up responses; 
the Maintainer Agent represents the evaluated LLM, explores the repository, and answers the user across multiple turns;
the Judge Agent evaluates the resulting conversation according to issue-specific satisfaction conditions and technical correctness. 
We set the maximum conversation length to 10 turns.
In our CAB experiments, both the User Agent and the Judge Agent use Gemini-2.5-Flash, following the backbone setting adopted in LoCoEval.
Before each interaction, the evaluation framework clones the target repository and checks out the commit specified by the dataset, after which the Maintainer Agent accesses the fixed repository snapshot through read-only code inspection and search tools. 
Among the 274 issues in CAB Verified v3, six could not be initialized due to unavailable repositories or cloning failures; therefore, all methods are evaluated on the same remaining 268 issues.

\subsection{Metrics}
\label{sec:evaluation_metrics}

We adopt different metrics following the two benchmarks.

\subsubsection{LoCoEval}
LoCoEval uses Pass@1 \cite{chen2021evaluating} as the primary effectiveness metric for repository-level function generation. A task is considered successfully completed if the generated code passes its execution-based test suite. Pass@1 denotes the percentage of successfully completed tasks.

We also report the context compression ratio (CR) following LoCoEval, which measures the reduction in total prompt-context consumption over the LLM calls involved, relative to the Full baseline. For a context-management method $M$, CR is calculated as:
\begin{equation}
    \mathrm{CR}(M)
    =
    \frac{\mathrm{InputTokens}(\mathrm{Full})}
         {\mathrm{InputTokens}(M)},
\end{equation}
where $\mathrm{InputTokens}(\cdot)$ denotes the cumulative number of input tokens consumed by all LLM calls under the same experimental setting. A larger CR indicates lower prompt-context consumption, with $\mathrm{CR}(\mathrm{Full}) = 1$.

CR characterizes context compression rather than the complete inference cost of a context-management method. We therefore additionally report end-to-end LLM token consumption, defined as the total input and output tokens consumed by response-generation and context-management LLM calls, excluding calls for mock-user simulation and judging. 
For \ourmethod{}, we further decompose this end-to-end token consumption into \emph{Turn Generation}, \emph{Task Generation}, \emph{Memory Sketching}, and \emph{Memory Update} to characterize the cost distribution across different stages.

We measure how well the retrieved context covers the ground-truth (GT) information items required by each function generation task. Let $G_i$ denote the original GT item set for task-run $i$, and let $s_{ig}^{v} \in \{0,1\}$ indicate whether context view $v$ fully supports item $g$. We consider three context views: the original-query context, the dual-query context, and their combination. GT-item coverage is calculated as:
\begin{equation}
    \operatorname{Coverage}(v)
    =
    \frac{1}{N}
    \sum_{i=1}^{N}
    \frac{\sum_{g \in G_i} s_{ig}^{v}}{|G_i|},
    \label{eq:gt_item_coverage}
\end{equation}
where $N$ is the number of evaluated task-runs, each corresponding to a function generation task under a particular backbone and task setting. Coverage is therefore macro-averaged over task-runs, giving each task-run equal weight regardless of its number of GT items. We further characterize coverage complementarity by dividing the GT items into four mutually exclusive categories: supported by the original-query context only, by both contexts independently, by the dual-query context only, or by neither. Each category's proportion is calculated within each task-run and then averaged across the same evaluation cohort. Combined-context coverage is evaluated independently on the merged context, rather than derived as the logical union of the two branch-level judgments. Since each context view is assessed in a separate LLM call, the combined score reflects an independent empirical judgment of the merged evidence and may therefore differ from the union implied by the two branch-level results.

\subsubsection{CodeAssistBench}
For CodeAssistBench, we follow its judge-based correctness evaluation. 
% After each interaction,
After each issue-level conversation concludes, 
the Judge Agent assigns one of three mutually exclusive labels---\emph{Correct}, \emph{Partially Correct}, or \emph{Incorrect}---based on the technical correctness of the Maintainer Agent's responses, their satisfaction of issue-specific user conditions, and, when applicable, execution results. \emph{Correct} indicates that the responses are technically accurate and fully satisfy the user conditions, whereas \emph{Partially Correct} indicates minor technical flaws or partial satisfaction of the conditions. We report the proportions of evaluated issues labeled \emph{Correct} and \emph{Partially Correct}; all remaining evaluated issues are classified as \emph{Incorrect}.

The Judge Agent additionally classifies response verbosity as \emph{Appropriate}, \emph{Verbose}, or \emph{Terse}, indicating whether the amount of information provided is suitable, excessive, or insufficient, respectively. 
We report the proportion of evaluated issues assigned to each category.
We also report the average number of interaction turns per issue (\emph{Avg. Turns}), where each turn corresponds to one response from the Maintainer Agent.

We distinguish between \emph{Maintainer tokens} and \emph{Memory tokens}. Maintainer tokens include all input and output tokens consumed by the Maintainer Agent, whereas Memory tokens capture the additional LLM token consumption introduced by an external memory mechanism. Their sum is reported as \emph{Combined tokens}. Let $N$ denote the number of evaluated issues, $r_i$ the number of interaction turns for issue $i$, and $T_i^x$ the token consumption of category $x$ for that issue, where $x \in \{\mathrm{Maintainer}, \mathrm{Memory}, \mathrm{Combined}\}$ and $T_i^{\mathrm{Combined}} = T_i^{\mathrm{Maintainer}} + T_i^{\mathrm{Memory}}$. The per-issue and per-turn token costs are calculated as:
\begin{equation}
    \mathrm{Tok./Issue}^{x}
    =
    \frac{1}{N}
    \sum_{i=1}^{N} T_i^x,
    \label{eq:tokens_per_issue}
\end{equation}
\begin{equation}
    \mathrm{Tok./Turn}^{x}
    =
    \frac{1}{N}
    \sum_{i=1}^{N}
    \frac{T_i^x}{r_i}.
    \label{eq:tokens_per_turn}
\end{equation}
Thus, $\mathrm{Tok./Turn}$ is macro-averaged over issues: the token cost per turn is first calculated separately for each issue and then averaged across all issues. This gives every issue equal weight regardless of its interaction length.

\subsection{Baselines}
\label{sec:baselines}

We compare \ourmethod{} with different memory settings.

\begin{itemize}
    \item \textbf{Full} retains the complete conversational history and truncates it at the turn level only when the context exceeds the model's context-window limit, without performing additional context selection or memory management.
    \item \textbf{Vanilla RAG} retrieves the top-$k$ user queries from the ongoing conversation history that are most similar to the current query and prepends the corresponding interaction turns as context. We set $k=5$ following LoCoEval.
    \item \textbf{MemGPT}~\cite{packer2023memgpt} adopts an OS-inspired virtual context management mechanism that manages information across hierarchical memory tiers.
    \item \textbf{LD-Agent}~\cite{li2025hello} organizes long-term conversational information through event-based memory and persona modeling, together with topic-oriented retrieval.
    \item \textbf{Mem0}~\cite{chhikara2025mem0} maintains persistent conversational memories through dynamic extraction, integration, and retrieval.
    \item \textbf{Mem0$^\mathcal{R}$}~\cite{liu2026scalable}, proposed together with LoCoEval, extends Mem0 specifically for repository-oriented long-horizon conversations. Instead of storing purely textual memories, Mem0R associates textual descriptions with the locations of relevant repository artifacts and uses these associations to retrieve the corresponding repository content when recalling memories. It therefore provides the most direct prior comparison for ReConMem, allowing us to examine ReConMem against an existing memory system that explicitly associates conversational information with repository artifacts.
\end{itemize}

Together, these methods cover different designs for explicit long-term memory management and allow us to examine how general-purpose memory paradigms transfer to repository-oriented conversations.

Because LoCoEval and CodeAssistBench adopt different interaction and evaluation frameworks, we adopt different comparisons. 

For \textbf{LoCoEval}, we compare \ourmethod{} with all six baselines.
For Full, Vanilla RAG, MemGPT, LD-Agent, and Mem0, we follow the LoCoEval protocol and provide repository access through a unified retrieval adaptation: for each user query, the 10 most similar repository functions are retrieved and supplied as additional context.
This avoids comparing repository-aware methods against general-purpose baselines that lack access to repository information and allows the evaluation to focus more directly on their context-management mechanisms.

For \textbf{CodeAssistBench}, we compare \ourmethod{} with three baselines that represent different context-management strategies within the same Maintainer Agent framework: \textbf{Full}, \textbf{Mem0}~\cite{chhikara2025mem0}, and \textbf{Strands Agents 1.45.0}\footnote{\url{https://strandsagents.com/}}, %a baseline specific to this benchmark 
a host-agent baseline included in our CAB experiments
that uses the native context-handling mechanism of the Strands-based Maintainer Agent without introducing an external memory system. It serves as the host-agent baseline and measures the performance achievable with the original agent context mechanism.

\subsection{Implementation Details}
\label{sec:evaluation_details}

Within each benchmark, all compared methods use the same response-generation prompt and repository-access configuration.
The temperature of all LLMs is set to 0 to reduce variability caused by probabilistic sampling.

Across both benchmarks, we implement \ourmethod{} with \textit{Tree-sitter}\footnote{\url{https://tree-sitter.github.io}} to parse code repositories and construct repository-side code indices, and \textit{all-MiniLM-L6-v2} \cite{wang2020minilm} with 384-dimensional embeddings to compute similarities for textual memories, code drafts, code entities, and repository retrieval.
The LLM calls involved in memory sketching and memory updating use the same backbone model as response generation.
The similarity threshold for code-draft-based query grounding is set to 0.35.

For LoCoEval, our experiments use three advanced backbone LLMs:
\textit{GPT-5 mini} \cite{openai2025gpt5mini}, \textit{DeepSeek-V3.2} \cite{liu2025deepseek32}, and \textit{Qwen3-235B}\footnote{We use its \texttt{qwen3-235b-a22b-instruct-2507} snapshot.} \cite{yang2025qwen3}, which have maximum context-window lengths of 400K, 128K, and 256K tokens, respectively.
These LLMs have broad user adoption and strong cost-effectiveness, and they cover both open-source and closed-source models.
For retrieval hyperparameters, $k_1$, $k_2$, $k_{31}$, and $k_{32}$ are set to 3, 6, 3, and 3 for \textit{DeepSeek-V3.2}, respectively; for the other two models, the values are 2, 4, 2, and 2.

To integrate \ourmethod{} into LoCoEval, we use it to manage the context provided to the LLM.
The preparation dialogue is processed sequentially, with each completed preparation turn triggering memory sketching and integration.
Before generating each response, \ourmethod{} retrieves relevant conversational memories and repository context.
For the final function-generation tasks, the accumulated memory state is queried in a read-only manner, preventing independently evaluated tasks from affecting one another.
We use all 128 LoCoEval conversation samples and evaluate their 320 function-generation tasks following the original conversation-generation and execution-based evaluation protocols.

In CAB, we use \textit{Qwen3-235B} as the backbone of the Maintainer Agent for all compared methods, using the same snapshot as in LoCoEval.
\ourmethod{} serves as the cross-turn context-management layer of the Maintainer Agent.
Because no interaction memory is available initially, the first Maintainer response is generated without memory recall, after which the completed interaction is incorporated into memory.
For each subsequent user follow-up, \ourmethod{} retrieves relevant conversational memories and repository context to augment the Maintainer Agent's input, and the memory state is updated after the response is completed.
\ourmethod{} is applied only to the Maintainer Agent; the User and Judge Agent configurations are held fixed across all compared methods.
We directly reuse the core memory implementation and the Qwen3-235B hyperparameter configuration from LoCoEval without additional tuning on CodeAssistBench.
Following the CodeAssistBench protocol, each conversation terminates when the user indicates that the issue has been resolved or when the maximum of 10 turns is reached.

\section{Results and Analysis}

\subsection{RQ1: Effectiveness}
\label{sec:rq1}

\begin{table*}[t]
\centering
\setlength{\abovecaptionskip}{0.1cm}
\caption{Experimental results of different methods and backbone models.
The values with arrows indicate the absolute differences in Pass@1
(in percentage points, pp) between \ourmethod{} and Mem0$^\mathcal{R}$, the strongest baseline in overall performance under the corresponding task setting.
CR denotes the prompt-token-based context compression ratio following LoCoEval.}
\label{tab:rq1_result_transposed}
\resizebox{\linewidth}{!}{
\begin{tabular}{l|cc|cc|cc|cc|cc|cc|cc}
\toprule
\multirow{2}{*}{\textbf{Model}} 
& \multicolumn{2}{|c|}{\textbf{\textit{Full}}} 
& \multicolumn{2}{c|}{\textbf{\textit{Vanilla RAG}}} 
& \multicolumn{2}{c|}{\textbf{MemGPT}} 
& \multicolumn{2}{c|}{\textbf{LD-Agent}} 
& \multicolumn{2}{c|}{\textbf{Mem0}} 
& \multicolumn{2}{c|}{\textbf{Mem0$^\mathcal{R}$}} 
& \multicolumn{2}{c}{\cellcolor{lightGray} \textbf{\ourmethod{}}}\\
\cmidrule(l{0.6em}){2-15}
& \textit{Pass@1} & \textit{CR} 
& \textit{Pass@1} & \textit{CR} 
& \textit{Pass@1} & \textit{CR} 
& \textit{Pass@1} & \textit{CR} 
& \textit{Pass@1} & \textit{CR} 
& \textit{Pass@1} & \textit{CR} 
& \cellcolor{lightGray} \textit{Pass@1} & \cellcolor{lightGray} \textit{CR} \\
\midrule

\multicolumn{15}{c}{\textbf{\textit{Single-hop}}} \\
\midrule

\textit{GPT-5 mini} 
& 35.000 & $1.0 \times$ 
& \textbf{\underline{40.625}} & $5.7 \times$ 
& 6.250 & $1.0 \times$ 
& 8.750 & $9.8 \times$ 
& 30.625 & $5.3 \times$ 
& 37.500 & $2.8 \times$
& \cellcolor{lightGray} 38.125~{\footnotesize\color{deepGreen} $\uparrow$ 0.625} & \cellcolor{lightGray} $2.9 \times$\\

\textit{DeepSeek-V3.2}
& 8.750 & $1.0 \times$ 
& 25.625 & $5.1 \times$ 
& 8.750 & $1.2 \times$ 
& 6.875 & $15.0 \times$ 
& 5.625 & $15.6 \times$ 
& \underline{31.875} & $5.1 \times$ 
& \cellcolor{lightGray} \textbf{41.875}~{\footnotesize\color{deepGreen} $\uparrow$ 10.000} & \cellcolor{lightGray} $5.0 \times$ \\

\textit{Qwen3-235B}
& 13.750 & $1.0 \times$ 
& 16.875 & $5.6 \times$ 
& 18.125 & $1.3 \times$ 
& 4.375 & $13.1 \times$ 
& 5.000 & $15.4 \times$ 
& \underline{24.375} & $6.1 \times$ 
& \cellcolor{lightGray} \textbf{38.125}~{\footnotesize\color{deepGreen} $\uparrow$ 13.750} & \cellcolor{lightGray} $5.6 \times$ \\

\midrule

\textbf{Overall}
& 19.167 & $1.0 \times$ 
& 27.708 & $5.4 \times$ 
& 11.042 & $1.2 \times$ 
& 6.667 & $12.6 \times$ 
& 13.750 & $10.1 \times$ 
& \underline{31.250} & $4.4 \times$
& \cellcolor{lightGray} \textbf{39.167}~{\footnotesize\color{deepGreen} $\uparrow$ 7.917} & \cellcolor{lightGray} $4.3 \times$ \\

\midrule
\multicolumn{15}{c}{\textbf{\textit{Multi-hop}}} \\
\midrule

\textit{GPT-5 mini} 
& 35.625 & $1.0 \times$ 
& 36.875 & $6.4 \times$ 
& 12.500 & $1.0 \times$ 
& 8.125 & $10.4 \times$ 
& 30.625 & $5.6 \times$ 
& \textbf{\underline{42.500}} & $3.2 \times$
& \cellcolor{lightGray} 40.625~{\footnotesize\color{deepGray} $\downarrow$ 1.875} & \cellcolor{lightGray} $3.1 \times$ \\

\textit{DeepSeek-V3.2} 
& 11.875 & $1.0 \times$ 
& 26.875 & $6.1 \times$ 
& 14.375 & $1.1 \times$ 
& 7.500 & $15.5 \times$ 
& 7.500 & $15.8 \times$ 
& \underline{37.500} & $5.4 \times$
& \cellcolor{lightGray} \textbf{45.625}~{\footnotesize\color{deepGreen} $\uparrow$ 8.125} & \cellcolor{lightGray} $5.0 \times$ \\

\textit{Qwen3-235B} 
& 18.750 & $1.0 \times$ 
& 18.125 & $5.8 \times$ 
& 21.250 & $1.1 \times$ 
& 6.250 & $12.9 \times$ 
& 9.375 & $15.8 \times$ 
& \underline{25.625} & $6.4 \times$
& \cellcolor{lightGray} \textbf{41.250}~{\footnotesize\color{deepGreen} $\uparrow$ 15.625} & \cellcolor{lightGray} $5.8 \times$ \\

\midrule

\textbf{Overall}
& 22.083 & $1.0 \times$ 
& 27.291 & $6.1 \times$ 
& 16.042 & $1.1 \times$ 
& 7.292 & $12.8 \times$ 
& 15.833 & $10.4 \times$ 
& \underline{35.208} & $4.7 \times$
& \cellcolor{lightGray} \textbf{42.500}~{\footnotesize\color{deepGreen} $\uparrow$ 7.292} & \cellcolor{lightGray} $4.4 \times$ \\

\bottomrule
\end{tabular}
}
% \vspace{-0.3cm}
\end{table*}

In controlled long-horizon context management scenarios, \ourmethod{} improves code generation effectiveness and context utilization compared with existing baselines.
As shown in Table~\ref{tab:rq1_result_transposed}, \ourmethod{} achieves the highest overall Pass@1 on both the single-hop and multi-hop subsets of LoCoEval across all evaluated backbone models. 
Compared with the overall strongest baseline Mem0$^\mathcal{R}$, \ourmethod{} improves the overall Pass@1 by 7.917 and 7.292 percentage points on the single-hop and multi-hop subsets, respectively, while maintaining a comparable context compression ratio.

The improvements are particularly pronounced on \textit{DeepSeek-V3.2} and \textit{Qwen3-235B}. 
Compared with Mem0$^\mathcal{R}$, \ourmethod{} improves Pass@1 by 10.000 and 13.750 percentage points on the single-hop subset, and by 8.125 and 15.625 percentage points on the multi-hop subset, respectively. 
In contrast, on GPT-5 mini, \ourmethod{} achieves competitive Pass@1 performance but does not consistently outperform the strongest baseline across both task settings.
This suggests that the benefits of explicit context management are not identical across different LLMs and are related to the backbone models' inherent capability in utilizing contextual information.

Notably, GPT-5 mini has the largest context window among the three evaluated backbone LLMs. 
Compared with \textit{DeepSeek-V3.2} and \textit{Qwen3-235B}, its Full History and Vanilla RAG baselines already achieve relatively strong Pass@1 performance, leaving less room for additional improvements from explicit memory management. 
This indicates that when backbone models are less effective at handling accumulated historical information or selecting task-relevant contexts, structured memory organization and targeted context retrieval can better compensate for these limitations and bring more substantial performance gains.

% \subsubsection{CodeAssistBench}
\begin{table}[t]
    \centering
    \small
    \setlength{\abovecaptionskip}{0.1cm}
    \caption{Performance comparison on CodeAssistBench.
Correctness and interaction-quality metrics are reported as percentages,
and Avg. Turns denotes the average number of conversation turns per issue.}
    \label{tab:cab_main_results}
    \begin{tabular}{lrrrr}
        \toprule
        \textbf{Metric}
        & \textbf{Native Strands}
        & \textbf{Full History}
        & \textbf{Mem0}
        & \cellcolor{lightGray} \textbf{\ourmethod{}} \\
        \midrule
    
        Correct (\%)
        & 19.78
        & 23.51
        & 19.40
        & \cellcolor{lightGray} \textbf{27.61} \\
    
        Partially Correct (\%)
        & 31.72
        & 36.94
        & 32.84
        & \cellcolor{lightGray} \textbf{44.03} \\
    
        \midrule
    
        Appropriate (\%) $\uparrow$
        & 60.45
        & 60.45
        & 62.31
        & \cellcolor{lightGray} \textbf{67.54} \\
    
        Verbose (\%) $\downarrow$
        & 27.24
        & 25.37
        & 30.60
        & \cellcolor{lightGray} \textbf{27.24} \\
    
        Terse (\%) $\downarrow$
        & 12.31
        & 14.18
        & 7.09
        & \cellcolor{lightGray} \textbf{5.22} \\
    
        \midrule
    
        Avg. Turns 
        & 4.30
        & 4.03
        & 4.18
        & \cellcolor{lightGray} \textbf{3.84} \\
    
        \bottomrule
    \end{tabular}
\end{table}

In realistic repository-oriented interaction scenarios, \ourmethod{} improves both task correctness and interaction quality compared with existing baselines.
As shown in Table~\ref{tab:cab_main_results}, Full History already achieves relatively strong performance by directly utilizing complete conversation histories, with Correct and Partially Correct rates of 23.51\% and 36.94\%, respectively.
Nevertheless, \ourmethod{} further improves these two metrics to 27.61\% and 44.03\%, achieving gains of 4.10 and 7.09 percentage points, respectively.
These results indicate that, in repository-oriented multi-turn interactions, effectively managing and utilizing contextual information is more beneficial than simply retaining the complete conversation history.
By organizing conversational memories and repository code information through structured representations and selecting relevant contexts according to the current request, \ourmethod{} enables the agent to better identify and utilize task-related information, thereby improving problem-solving performance.

In contrast, the general-purpose memory method Mem0 provides limited benefits in this repository-oriented setting.
Its Correct and Partially Correct rates are 19.40\% and 32.84\%, respectively, which are even lower than those achieved by Full History without explicit memory management.
This further suggests that context management for software engineering tasks requires modeling the structural associations between conversational information and repository code, while simply extracting and storing textual memories from historical interactions is insufficient to effectively improve task performance.

Beyond correctness metrics, Table~\ref{tab:cab_main_results} shows that \ourmethod{} achieves the highest Appropriate rate of 67.54\% among all compared methods.
Meanwhile, it requires the fewest average interaction turns, with only 3.84 turns compared with 4.30, 4.03, and 4.18 turns for Native Strands, Full History, and Mem0, respectively.
These results indicate that the improvement achieved by \ourmethod{} does not rely on generating longer responses or requiring additional interaction rounds.
Instead, by retrieving and utilizing more relevant context, \ourmethod{} enables the agent to provide responses that better satisfy user requirements while reducing unnecessary clarification and correction interactions.

\begin{custommdframed}
\textbf{\textit{Answer to RQ1:}}
\ourmethod{} demonstrates strong overall effectiveness across both controlled long-horizon and realistic repository-oriented conversational settings. It achieves the highest overall Pass@1 on both LoCoEval subsets while maintaining a comparable context compression ratio, and achieves the highest Correct, Partially Correct, and Appropriate rates on CodeAssistBench with the fewest average interaction turns.
\end{custommdframed}

%%%%%%%%%%%%%%%%%%%%%%%%%%%%%%%%
\subsection{RQ2: Efficiency}
\label{sec:rq4}
Beyond the context compression ratio reported in Table~\ref{tab:rq1_result_transposed}, we further examine the end-to-end LLM token consumption of \ourmethod{} on LoCoEval.
We first examine the context-management cost of \ourmethod{} on LoCoEval. As shown in Table~\ref{tab:rq4_token_cost}, \ourmethod{} substantially reduces token consumption compared with Full across all backbone models and task settings, with an approximately 55\%--79\% reduction in the average token consumption per conversation. Meanwhile, its token consumption remains comparable to Mem0$^\mathcal{R}$, indicating that the performance gains observed in RQ1 are not obtained by substantially increasing context cost.

\begin{table*}[t]
\centering
\setlength{\abovecaptionskip}{0.1cm}
\caption{Average end-to-end LLM token consumption per conversation across different methods and backbone models on LoCoEval. The Average row reports the mean across the three backbone models.}
\label{tab:rq4_token_cost}

\small
\setlength{\tabcolsep}{6pt}

\begin{tabular}{l|rrrrrrr}
\toprule
\textbf{Model} 
& \textbf{\textit{Full}} 
& \textbf{\textit{Vanilla RAG}} 
& \textbf{\textit{MemGPT}} 
& \textbf{\textit{LD-Agent}} 
& \textbf{\textit{Mem0}} 
& \textbf{\textit{Mem0$^\mathcal{R}$}} 
& \cellcolor{lightGray}\textbf{\ourmethod{}} \\
\midrule

\multicolumn{8}{c}{\textbf{\textit{Single-hop}}} \\
\midrule
\textit{GPT-5 mini} 
& 3,388K
& 694K
& 3,611K
& 388K
& 1,075K
& 1,584K
& \cellcolor{lightGray} 1,534K \\

\textit{DeepSeek-V3.2}
& 4,592K
& 1,028K
& 3,888K
& 309K
& 385K
& 1,072K
& \cellcolor{lightGray} 1,090K \\

\textit{Qwen3-235B}
& 4,162K
& 831K
& 3,154K
& 328K
& 369K
& 859K
& \cellcolor{lightGray} 907K \\

\midrule

\textbf{Average}
& \textbf{4,047K}
& \textbf{851K}
& \textbf{3,551K}
& \textbf{342K}
& \textbf{610K}
& \textbf{1,171K}
& \cellcolor{lightGray} \textbf{1,177K} \\

\midrule
\multicolumn{8}{c}{\textbf{\textit{Multi-hop}}} \\
\midrule

\textit{GPT-5 mini} 
& 3,275K
& 607K
& 3,455K
& 358K
& 983K
& 1,368K
& \cellcolor{lightGray} 1,420K \\

\textit{DeepSeek-V3.2} 
& 4,303K
& 807K
& 3,887K
& 282K
& 351K
& 930K
& \cellcolor{lightGray} 1,018K \\

\textit{Qwen3-235B} 
& 3,902K
& 753K
& 3,348K
& 315K
& 331K
& 762K
& \cellcolor{lightGray} 826K \\

\midrule

\textbf{Average}
& \textbf{3,827K}
& \textbf{722K}
& \textbf{3,564K}
& \textbf{318K}
& \textbf{555K}
& \textbf{1,020K}
& \cellcolor{lightGray} \textbf{1,088K} \\

\bottomrule
\end{tabular}

% \vspace{-0.3cm}
\end{table*}

Table~\ref{tab:rq4_reconmem_breakdown} further decomposes the cost introduced by \ourmethod{}. Memory Sketching and Memory Update together account for 38.33\% and 39.35\% of the total token consumption on the single-hop and multi-hop subsets, respectively, with Memory Update constituting the larger portion. Thus, explicit memory management itself introduces non-negligible overhead. The efficiency of \ourmethod{} therefore depends on whether the cost of extracting and maintaining useful context can be offset by reducing the irrelevant information subsequently supplied to the backbone model. Notably, this stage-wise cost distribution remains similar across different backbone models and task settings.

\begin{table*}[t]
\centering
\setlength{\abovecaptionskip}{0.1cm}
\caption{Token consumption breakdown of \ourmethod{} across different backbone models on LoCoEval. 
Tok./Conv. denotes the average number of input and output tokens consumed per conversation, while Share denotes the proportion of tokens consumed by each stage relative to the total token consumption of \ourmethod{}.}
\label{tab:rq4_reconmem_breakdown}

\small
\setlength{\tabcolsep}{4.5pt}

\begin{tabular}{l|cc|cc|cc|cc|c}
\toprule
\multirow{2}{*}{\textbf{Model}}
& \multicolumn{2}{c|}{\textbf{Turn Generation}}
& \multicolumn{2}{c|}{\textbf{Task Generation}}
& \multicolumn{2}{c|}{\textbf{Memory Sketching}}
& \multicolumn{2}{c|}{\textbf{Memory Update}}
& \multicolumn{1}{c}{\textbf{Total}} \\
\cmidrule(lr{0.5em}){2-10}
& \textit{Tok./Conv.} & \textit{Share (\%)}
& \textit{Tok./Conv.} & \textit{Share (\%)}
& \textit{Tok./Conv.} & \textit{Share (\%)}
& \textit{Tok./Conv.} & \textit{Share (\%)}
& \textit{Tok./Conv.}  \\
\midrule

\multicolumn{10}{c}{\textbf{\textit{Single-hop}}} \\
\midrule
\textit{GPT-5 mini}
& 911K & 59.42
& 52K & 3.43
& 174K & 11.36
& 395K & 25.79
& 1,534K  \\

\textit{DeepSeek-V3.2}
& 639K & 58.62
& 30K & 2.78
& 117K & 10.81
& 302K & 27.79
& 1,090K  \\

\textit{Qwen3-235B}
& 526K & 57.99
& 18K & 2.04
& 114K & 12.64
& 248K & 27.34
& 907K  \\

\midrule

\rowcolor{lightGray}
\textbf{Average}
& \textbf{692K} & \textbf{58.80}
& \textbf{33K} & \textbf{2.87}
& \textbf{135K} & \textbf{11.52}
& \textbf{315K} & \textbf{26.81}
& \textbf{1,177K}  \\

\midrule
\multicolumn{10}{c}{\textbf{\textit{Multi-hop}}} \\
\midrule

\textit{GPT-5 mini}
& 837K & 58.97
& 44K & 3.16
& 167K & 11.81
& 370K & 26.07
& 1,420K  \\

\textit{DeepSeek-V3.2}
& 590K & 58.00
& 27K & 2.66
& 114K & 11.21
& 286K & 28.13
& 1,018K \\

\textit{Qwen3-235B}
& 464K & 56.19
& 15K & 1.92
& 109K & 13.22
& 237K & 28.67
& 826K  \\

\midrule

\rowcolor{lightGray}
\textbf{Average}
& \textbf{630K} & \textbf{57.96}
& \textbf{29K} & \textbf{2.69}
& \textbf{130K} & \textbf{11.98}
& \textbf{297K} & \textbf{27.37}
& \textbf{1,088K}  \\

\bottomrule
\end{tabular}

% \vspace{-0.3cm}
\end{table*}

\begin{figure}[t]
    \centering
    \setlength{\abovecaptionskip}{0.2cm}
    \includegraphics[width=0.65\linewidth]{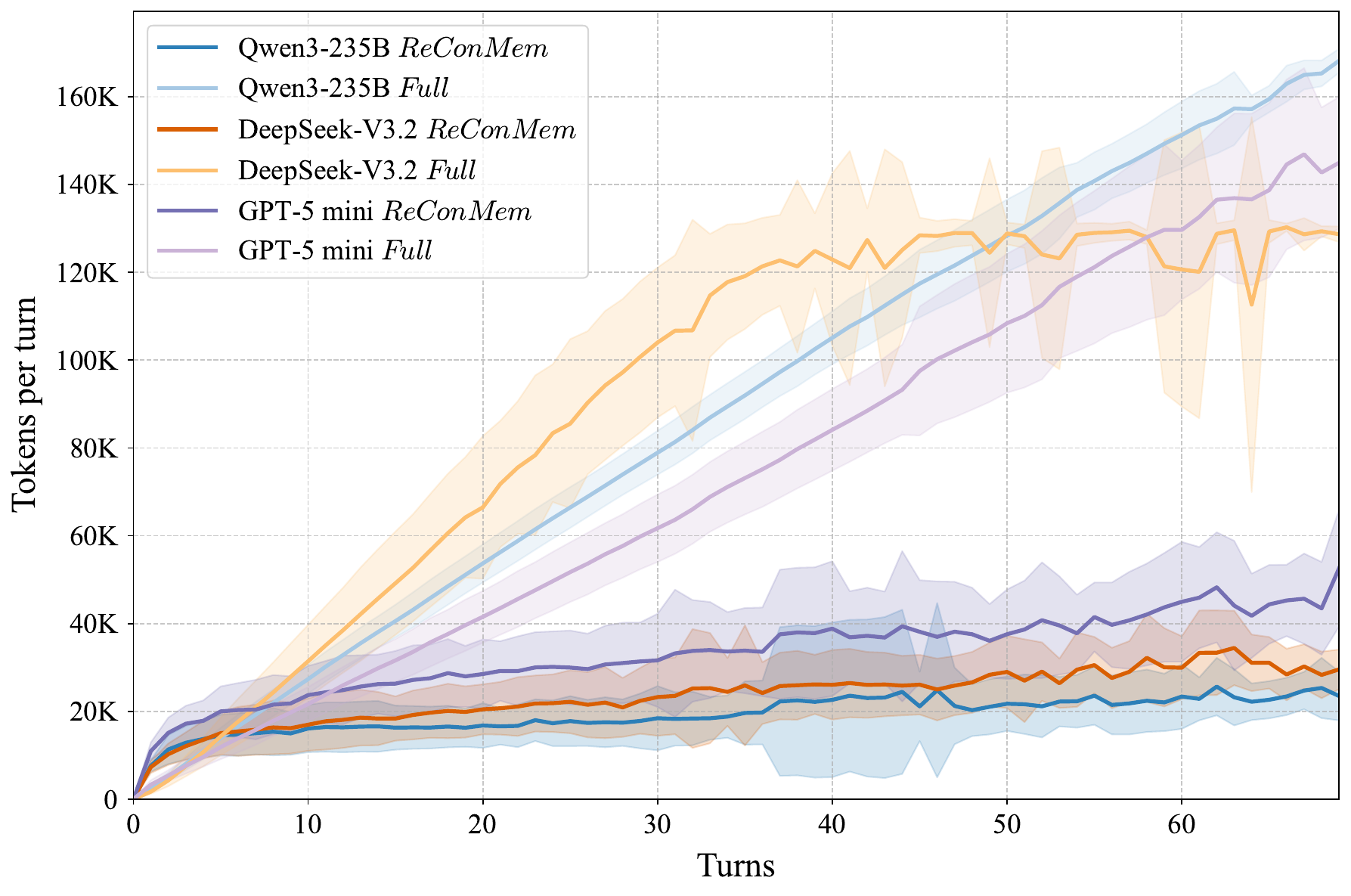}
    \caption{Per-turn token consumption of \ourmethod{} and \textit{Full} as conversation turns increase on LoCoEval.}
    \label{fig:tokens_per_turn}
    % \vspace{-0.3cm}
\end{figure}

Figure~\ref{fig:tokens_per_turn} further shows how this trade-off changes with conversation length. At the early stages of a conversation, the history is still short, and the additional memory-management operations can make \ourmethod{} more expensive than directly using the accumulated context. As the conversation progresses, however, the per-turn token consumption of Full continues to increase with the growing history, whereas \ourmethod{} grows substantially more slowly by selectively retaining and recalling relevant information. From turn 10 to turn 60, the number of turns increases by a factor of six, whereas the per-turn token consumption of \ourmethod{} increases by less than twofold. Therefore, the token-saving benefit of \ourmethod{} becomes increasingly pronounced as more historical information accumulates, because a larger amount of irrelevant or currently unnecessary context can be excluded from each turn.

The CodeAssistBench results provide a complementary short-horizon perspective on this behavior. Unlike LoCoEval, CAB conversations terminate dynamically when the User Agent considers the issue resolved or when the maximum of 10 turns is reached, and the evaluated methods consequently exhibit different interaction lengths. Therefore, Tok./Issue and Tok./Turn should be interpreted jointly rather than as isolated measures of context-management efficiency: the former reflects both per-turn cost and the number of turns required to complete an interaction, whereas the latter normalizes conversation length but is still measured over method-dependent interaction trajectories. In practice, CAB conversations are short, averaging only around four turns. Under this setting, the historical context has not yet accumulated sufficiently for the savings from selective context management to offset its memory-management overhead. Accordingly, as shown in Table~\ref{tab:cab_efficiency}, \ourmethod{} consumes 363,822.40 Combined Tok./Issue and 100,921.00 Combined Tok./Turn, compared with 325,622.92 and 84,972.27 for Full History, respectively. This result is consistent with the early-stage behavior observed in Figure~\ref{fig:tokens_per_turn}.

Nevertheless, the CAB results also show that the additional cost introduced by \ourmethod{} is effectively controlled. Its Memory Tok./Issue is 17,241.94, approximately 56\% lower than the 39,203.21 of Mem0, while its Combined Tok./Issue is approximately 10\% lower. Moreover, the higher token consumption of \ourmethod{} relative to Full History should be considered together with its interaction outcomes. As reported in RQ1, \ourmethod{} achieves the highest Correct (27.61\%), Partially Correct (44.03\%), and Appropriate (67.54\%) rates while requiring the fewest average turns (3.84). In particular, the higher Appropriate rate provides complementary evidence that its additional token consumption does not simply result from unnecessarily verbose responses. Taken together, these results suggest that, even before explicit token savings emerge in short conversations, selective context management can still improve the effectiveness of repository-oriented interactions by focusing the model on useful conversational and repository context.

\begin{table}[t]
    \centering
    \setlength{\abovecaptionskip}{0.1cm}
    \caption{Token consumption of different context-management methods on CodeAssistBench. Maintainer tokens include all input and output tokens consumed by the Maintainer Agent, while memory tokens include the additional LLM token consumption introduced by the external memory mechanism. Combined tokens are the sum of the two. Tok./Turn is macro-averaged over issues.}
    \label{tab:cab_efficiency}

    \begin{tabular}{lrrrr}
        \toprule
        \textbf{Metric}
        & \textbf{Native Strands}
        & \textbf{Full History}
        & \textbf{Mem0}
        & \textbf{\ourmethod{}} \\
        \midrule

        Maintainer Tok./Issue
        & 360,521.97
        & 325,622.92
        & 365,017.51
        & 346,580.46 \\

        Memory Tok./Issue
        & 0
        & 0
        & 39,203.21
        & 17,241.94 \\

        Combined Tok./Issue
        & 360,521.97
        & 325,622.92
        & 404,220.72
        & 363,822.40 \\

        \midrule

        Maintainer Tok./Turn
        & 96,825.50
        & 84,972.27
        & 94,146.95
        & 96,357.12 \\

        Memory Tok./Turn
        & 0
        & 0
        & 9,437.53
        & 4,563.88 \\

        Combined Tok./Turn
        & 96,825.50
        & 84,972.27
        & 103,584.48
        & 100,921.00 \\

        \bottomrule
    \end{tabular}
\end{table}

\begin{custommdframed}
\textbf{\textit{Answer to RQ2:}}
\ourmethod{} introduces additional memory-management overhead, but selectively extracting, consolidating, and recalling relevant context limits the growth of contextual cost as conversation history accumulates. Consequently, its token-efficiency advantage becomes increasingly evident in long-horizon conversations. In shorter interactions, this overhead may not yet be fully offset by token savings, but \ourmethod{} still maintains controlled memory cost and improves interaction effectiveness without relying on longer or excessively verbose conversations. Overall, \ourmethod{} supports both efficient long-horizon context management and effective context utilization in shorter repository-oriented interactions.
\end{custommdframed}

\subsection{RQ3: Ablation Study}
\label{sec:rq2}

To evaluate the effectiveness of each core component in \ourmethod{}, we conduct ablation studies on LoCoEval. We focus the ablation analysis on LoCoEval because its controlled long-horizon evaluation setting allows the effects of individual context-management components on generation accuracy and context efficiency to be examined more directly.
Specifically, we remove or modify key components as follows: 
\ding{182} \textbf{w/o GAE}: removing graph-based adjacency expansion and using similarity-based retrieval only; 
\ding{183} \textbf{w/o SMG}: removing direct retrieval on the structural memory graph, and instead allowing code entities to be obtained solely through expansion via inter-graph adjacency; 
\ding{184} \textbf{w/o QG}: disabling code-draft-based query grounding and directly using the original query for code-side retrieval; and
\ding{185} \textbf{w/o HCC}: disabling hierarchical code context compression and retaining full code snippets after retrieval. 
To control variables, we adjust the hyperparameters to keep the compression ratio under different ablation settings as consistent as possible with the reference value. All experiments use \textit{Qwen3-235B} as the backbone model.

The ablation results in Table \ref{tab:rq2_result} show that removing any key component leads to a decrease in Pass@1, indicating that each module contributes to the overall framework. Specifically, removing graph-based adjacency expansion (w/o GAE) decreases Pass@1 by 7.500 and 10.625 percentage points on the single-hop and multi-hop subsets, respectively. Removing direct retrieval on the structural memory graph (w/o SMG) results in decreases of 10.000 and 6.875 percentage points. These two settings cause the largest performance degradation among all ablations, while their changes in compression ratio remain comparatively limited, highlighting the importance of the memory graph structure and joint retrieval mechanism for generation accuracy.

In contrast, removing hierarchical code context compression (w/o HCC) mainly affects context efficiency. Its Pass@1 decreases by only 1.875 and 0.625 percentage points on the single-hop and multi-hop subsets, respectively, while the compression ratio decreases substantially by 29.60\% and 27.92\%. This demonstrates that HCC primarily contributes to controlling code-context consumption while preserving generation accuracy. In addition, disabling code-draft-based query grounding (w/o QG) reduces Pass@1 by 1.875 and 3.750 percentage points on the two subsets, respectively, while having little effect on the compression ratio. This result shows that the code-draft-based query grounding provides consistent additional gains in generation accuracy.
\begin{table}[t]
\centering
\setlength{\abovecaptionskip}{0.1cm}
\caption{Ablation study results on LoCoEval using Qwen3-235B.
The values with arrows indicate the absolute differences in Pass@1 (in percentage points, pp) 
compared with the complete \ourmethod{} configuration.}
\label{tab:rq2_result}

\small
\setlength{\tabcolsep}{4.5pt}

\begin{tabular}{l|ll|ll}
\toprule
\multirow{2}{*}{\textbf{Method}} 
& \multicolumn{2}{c|}{\textbf{\textit{Single-hop}}} 
& \multicolumn{2}{c}{\textbf{\textit{Multi-hop}}} \\
\cmidrule(lr{0.6em}){2-5}
& \textit{Pass@1} & \textit{CR} 
& \textit{Pass@1} & \textit{CR} \\
\midrule

\rowcolor{lightGray} {\textbf{\ourmethod{}}} 
& 38.125 
& $5.6 \times$ 
& 41.250 
& $5.8 \times$ \\
\midrule

w/o GAE 
& 30.625~{\footnotesize\color{deepRed} $\downarrow$ 7.500} 
& $6.2 \times$
& 30.625~{\footnotesize\color{deepRed} $\downarrow$ 10.625} 
& $5.4 \times$\\

w/o SMG 
& 28.125~{\footnotesize\color{deepRed} $\downarrow$ 10.000} 
& $5.2 \times$
& 34.375~{\footnotesize\color{deepRed} $\downarrow$ 6.875} 
& $5.3 \times$ \\

w/o QG
& 36.250~{\footnotesize\color{deepRed} $\downarrow$ 1.875} 
& $5.8 \times$
& 37.500~{\footnotesize\color{deepRed} $\downarrow$ 3.750} 
& $5.8 \times$\\

w/o HCC 
& 36.250~{\footnotesize\color{deepRed} $\downarrow$ 1.875} 
& $4.0 \times$
& 40.625~{\footnotesize\color{deepRed} $\downarrow$ 0.625} 
& $4.2 \times$ \\

\bottomrule
\end{tabular}

% \vspace{-0.3cm}
\end{table}

\begin{custommdframed}
\textbf{\textit{Answer to RQ3:}}
All four components contribute to \ourmethod{}, but in different ways. GAE and direct retrieval on the structural memory graph contribute most to generation accuracy, QG provides additional accuracy gains, and HCC primarily improves context efficiency.
\end{custommdframed}

\subsection{RQ4: Context Retrieval Quality}
\label{sec:rq3}
To examine the coverage and complementarity of the contexts retrieved by \ourmethod{}, 
we analyze 960 LoCoEval task-runs across the three backbone models and the single-hop and multi-hop settings, covering 37 repositories. 
For each task-run, we evaluate the original-query context, the dual-query context, and their merged context after removing exact duplicate snippets within each modality. 
Each context view is evaluated independently by \textit{Gemini-2.5-Flash} with temperature set to 0 and a fixed prompt containing eight synthetic examples. 
We use the ground-truth (GT) items that LoCoEval provides as evaluation targets. A GT item is considered supported only when the supplied context provides sufficient evidence for all its substantive requirements; partial support receives no credit, and neither the task query nor the reference code is treated as retrieved evidence. Positive judgments must reference valid evidence identifiers in the supplied context, and structured outputs are checked automatically for item completeness and evidence-identifier validity.
We estimate percentile-based 95\% confidence intervals using 2,000 repository-cluster bootstrap resamples.

\begin{figure}[t]
    \centering
    \includegraphics[width=\linewidth]
    {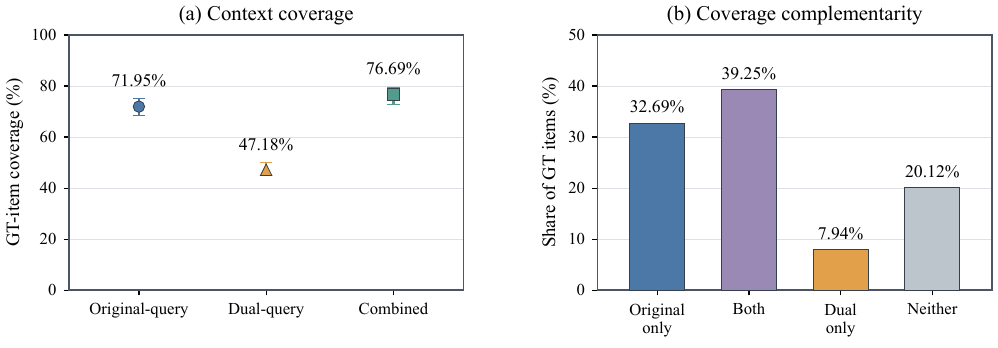}
    \caption{
        Context coverage and complementarity of \ourmethod{} on LoCoEval.
        (a) Mean GT-item coverage of the original-query, dual-query and combined contexts.
        (b) Proportions of GT items supported by the original-query context only, both contexts independently, the dual-query context only, or neither.
        All values are macro-averaged over task-runs.
        Error bars indicate 95\% repository-cluster bootstrap confidence intervals.
    }
    \label{fig:rq3_coverage}
\end{figure}

Figure~\ref{fig:rq3_coverage}(a) reports the GT-item coverage of the original-query, dual-query, and combined contexts, while Figure~\ref{fig:rq3_coverage}(b) characterizes the complementary contributions of the two retrieval branches.

As shown in Figure~\ref{fig:rq3_coverage}(a), the original-query context achieves an average coverage of 71.95\%, compared with 47.18\% for the dual-query context. The independently evaluated combined context achieves 76.69\%, exceeding the original-query context by 4.75 percentage points, with a 95\% confidence interval of [2.09, 6.83] percentage points for the increase.
These results indicate that the original-query branch provides broader coverage under the evaluated configuration, while incorporating the dual-query results increases the coverage of task requirements.

Figure~\ref{fig:rq3_coverage}(b) further characterizes the coverage contributed by the two branches. On average, 39.25\% of GT items are supported by both contexts independently, 32.69\% are supported only by the original-query context, and 7.94\% are supported only by the dual-query context. Thus, despite its lower standalone coverage, the dual-query branch provides task-relevant evidence absent from the original-query results. The substantial original-only coverage likewise shows that the two branches make distinct contributions and cannot fully substitute for one another. The remaining 20.12\% of items are unsupported by either context when evaluated separately.

One interpretation of this pattern concerns the different retrieval entry points. The original query starts from textual memories and more directly accesses descriptions of requirements, design decisions, and behavioral constraints. The dual query provides a code-oriented entry point that can lead to concrete interfaces, dependencies, and implementation behavior. Such information may cover fewer GT items while still providing specific references for implementing the corresponding requirements. 
The two retrieval routes may therefore emphasize different aspects of requirement coverage and implementation guidance, with the observed dual-only coverage providing direct evidence of their complementarity.

Moreover, both branches undergo graph-based expansion, and their output sizes depend on the initial retrieval budgets, accessible neighborhoods, and overlap among entities. They are not constrained to equal snippet counts or token budgets, and both can return textual and code content through cross-graph relations. The coverage difference therefore describes their contributions under the current retrieval configuration, rather than ranking the two modalities under an equal budget. The central observations are the exclusive information provided by each branch and the coverage improvement obtained from their combination.

\begin{figure}[t]
    \centering
    \setlength{\abovecaptionskip}{0.2cm}
    \includegraphics[width=\linewidth]{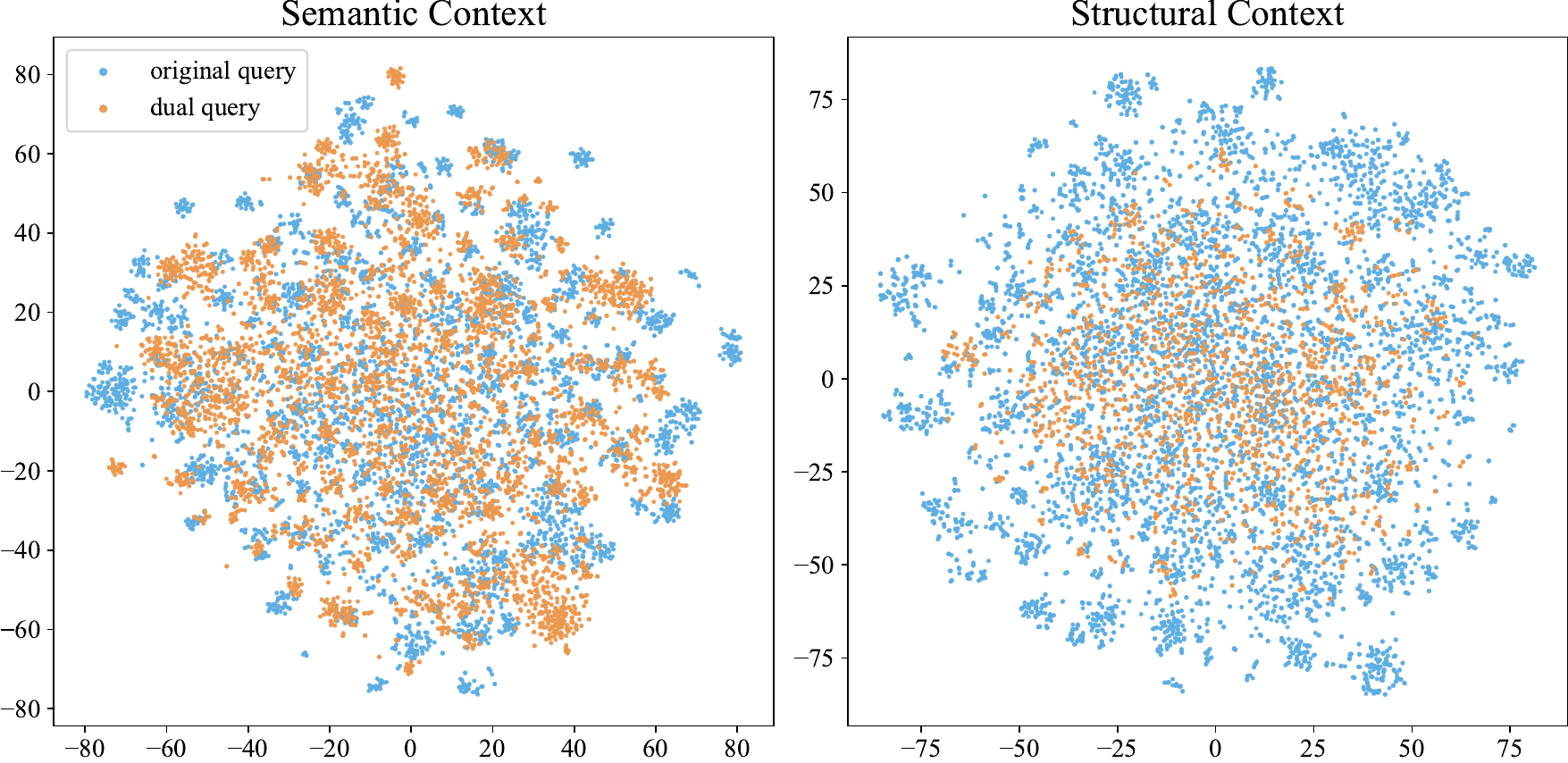}
    \caption{t-SNE visualization of retrieved semantic (left) and structural (right) contexts from the dual memory graph, obtained via the original query (blue) and dual query (orange). The visualization qualitatively reflects local distributional overlap and differences between the two retrieval routes.}
    \label{fig:memory_embeddings}
    % \vspace{-0.3cm}
\end{figure}

To illustrate the relationship between the retrieval results,
Figure~\ref{fig:memory_embeddings} presents t-SNE visualizations of the semantic and structural contexts. Points associated with the two query sources overlap but also exhibit differences in their local distributions, with some local groups more prominently represented by one source.
This pattern is consistent with the shared and exclusive coverage in Figure~\ref{fig:rq3_coverage}: the two branches retrieve some common information while also returning distinct context. The visualization provides a qualitative view of their distributions, while the GT-item evaluation establishes their complementary support for task requirements.

\begin{custommdframed}
\textbf{\textit{Answer to RQ4:}}
\ourmethod{}'s two retrieval branches provide asymmetric but complementary coverage of task-relevant information. The original-query branch achieves broader coverage, while the dual-query branch exclusively supports 7.94\% of GT items. Their combined context achieves 76.69\% average coverage, exceeding the original-query context by 4.75 percentage points and supporting the effectiveness of using both retrieval routes.
\end{custommdframed}

\section{Discussion}

\subsection{Performance under Varying Conversation Characteristics}
\label{sec:conversation_characteristics}

\begin{figure}[t]
    \centering
    \setlength{\abovecaptionskip}{0.1cm}
    \includegraphics[width=0.8\linewidth]{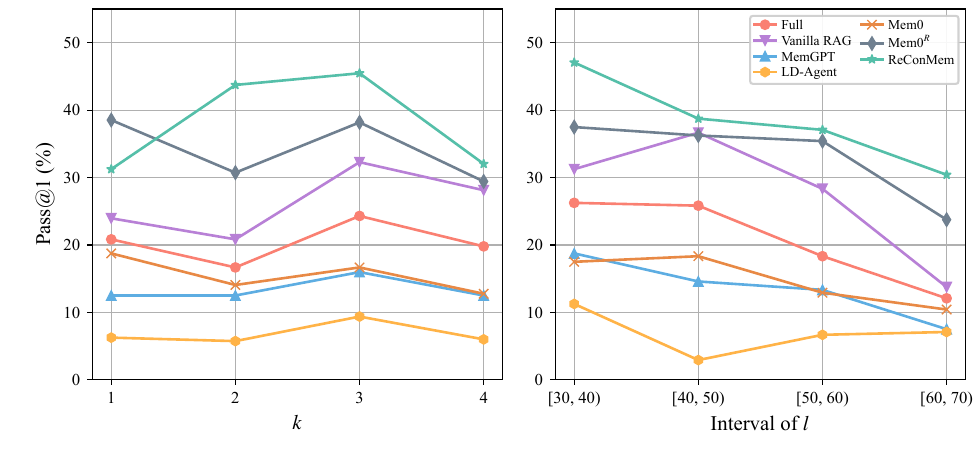}
    \caption{Trend of Pass@1 for each method, with respect to the number of tasks per conversation $k$ (left) and the number of turns per conversation $l$ (right).}
    \label{fig:k_and_l_analysis}
    \vspace{-0.3cm}
\end{figure}

To understand how different conversation characteristics affect context-management performance, we further analyze the experimental results of \ourmethod{} on LoCoEval. Specifically, we partition the LoCoEval conversations according to the number of code-generation tasks ($k$) and conversation length ($l$), and report the Pass@1 performance of different methods in Figure~\ref{fig:k_and_l_analysis}.

As shown in Figure~\ref{fig:k_and_l_analysis} (left), \ourmethod{} achieves the highest Pass@1 among all evaluated methods in the multi-task settings ($k=2,3,4$), whereas Mem0$^\mathcal{R}$ slightly outperforms \ourmethod{} in the single-task setting. This suggests that \ourmethod{} tends to exhibit a clearer advantage when a conversation involves multiple code-generation tasks and the relevant context may be distributed across different parts of the interaction. However, the performance of \ourmethod{} does not increase monotonically with the number of tasks. It achieves its best performance at $k=3$, while its Pass@1 decreases by 13.45 percentage points when $k$ further increases to 4. This indicates that increasing task complexity provides more opportunities to leverage historical context, while also introducing additional challenges caused by information competition and retrieval difficulty.

Conversation length exhibits a similar pattern. As shown in Figure~\ref{fig:k_and_l_analysis} (right), the performance of all methods decreases as conversations become longer, including that of \ourmethod{}. In comparison, the advantage of \ourmethod{} lies in maintaining stronger relative performance as conversational context accumulates. In the longest conversation interval $[60,70)$, both \ourmethod{} and the baselines exhibit a clear drop in Pass@1, but the decrease of \ourmethod{} is less pronounced than those of the baselines.

Overall, our results support a more restrained conclusion: by explicitly maintaining memories of conversational history and repository code, together with their associations, \ourmethod{} can effectively mitigate the retrieval difficulty and context interference caused by information accumulation, but it cannot make the model completely insensitive to increasing interaction complexity.

\subsection{Threats to Validity}

\textbf{Threats to internal validity} primarily arise from differences in context consumption across methods and the inherent variability of LLM-based processing.
On LoCoEval, we carefully tune the retrieval-related hyperparameters to keep the token usage of \ourmethod{} as comparable as possible to that of the strongest baseline, Mem0$^\mathcal{R}$.
However, because conversations evolve dynamically and different methods employ different context-management mechanisms, it is difficult to strictly control the final token consumption, making some discrepancies unavoidable.
Nevertheless, as discussed in \S\ref{sec:rq1}, the differences in context compression ratio are relatively limited compared with the observed performance gains.
In addition, although we set the temperature of all backbone LLMs to 0 to reduce variability caused by probabilistic sampling, model generation may still exhibit minor nondeterminism due to model-serving and hardware-level factors.

Another potential threat arises from the use of LLM-based judgments in the evaluation.
CodeAssistBench relies on a Judge Agent to assess correctness and interaction quality, while our context retrieval quality analysis also uses an LLM to determine whether the retrieved contexts sufficiently support the ground-truth information items.
Such judgments may introduce evaluation variance or model-specific bias.
To mitigate this threat, we follow the original CodeAssistBench evaluation protocol and keep the Judge Agent unchanged across all compared methods.
For the context retrieval quality evaluation, we first calibrate the LLM on a subset of examples, analogous to a training set.
We then freeze the calibrated prompt and evaluate its agreement on a separate set of examples, analogous to a validation set.
Full-scale annotation is conducted only after all of these consistency checks are passed.
Throughout this process, we use exactly the same evaluator, temperature, prompt, and decision criteria for all context views.
These controls improve evaluation consistency across different experimental conditions, although they cannot completely eliminate the uncertainty inherent in LLM-based evaluation.

\textbf{Threats to external validity} mainly concern the generalizability of our results.
Repository-oriented conversational context management remains a relatively emerging research setting, and currently available benchmarks cover only a limited range of development tasks and interaction patterns.
We evaluate \ourmethod{} on two complementary benchmarks, LoCoEval and CodeAssistBench.
Although this combination broadens the evaluation beyond a single benchmark, it still cannot cover all repository-oriented development scenarios, such as longer-term cross-session development processes and other forms of repository evolution.

Finally, due to computational and budget constraints, our evaluation covers only a limited set of backbone models.
LoCoEval is evaluated with three different LLMs, whereas CodeAssistBench uses Qwen3-235B as the Maintainer backbone under a fixed agent configuration.
Although the results across different LoCoEval backbones and the complementary CodeAssistBench setting exhibit consistent overall trends, the exact performance gains and efficiency trade-offs may vary across different models, repository distributions, and interaction settings.

\section{Related Work}

With the increasing application of large language models (LLMs) in multi-turn conversations and complex software development tasks \cite{he2025llm, chen2025empirical}, effectively leveraging historical context over long-horizon interactions has become a critical research challenge \cite{mei2025survey}. Due to the limitation of context windows, models are unable to directly process excessively long conversational sequences, which has motivated a growing body of work on external mechanisms to enhance context management.
Early approaches primarily construct such mechanisms based on simple retrieval-augmented generation (RAG), where historical conversations are stored and relevant segments are retrieved at query time to augment generation, such as RECOMP \cite{xu2024recomp} and COCOM \cite{rau2024context}.
While these methods are simple to implement and scalable, they typically manage context at a coarse granularity, such as turn-level, which can easily introduce a large amount of noisy information.

To address this limitation, subsequent work explores more fine-grained context management strategies.
The dominant paradigm, referred to as memory systems, periodically extracts conversational context into a set of atomic memory units, which are retrieved as background knowledge in each turn.
Early memory systems typically store and retrieve memory units independently \cite{zhong2024memorybank, packer2023memgpt, li2025memos, tan2025prospect, li2025hello}, overlooking the semantic relationships among them, which leads to issues such as fragmented retrieval results and insufficient accuracy \cite{Amem, zhang2025bridging}.
To address this, some studies \cite{chhikara2025mem0, Amem, rasmussen2025zep, huang2026licomemory, wu2025sgmem, tang2026mnemis} further propose leveraging graphs to model relationships among memories, enabling more coherent retrieval and improved coverage. 

Building upon these developments, as code LLMs extend toward repository-level software engineering tasks \cite{bairi2024codeplan, ma2025alibaba, yang2024swe, zhang2024codeagent}, recent research has begun to address scenarios involving both code repositories and multi-turn conversations. Compared to single-turn or short-context code generation, such tasks require continuously leveraging historical interactions while reasoning over structured code context within repositories, posing higher demands on memory modeling.
In terms of evaluation data, LoCoEval \cite{liu2026scalable} is the first benchmark to systematically define long-horizon conversational context management for repository-level settings. It constructs multi-turn interaction processes and evaluates model performance on repository-level tasks based on conversational information, thereby assessing the model’s ability to accumulate and utilize long-horizon context. Experimental results show that existing general-purpose memory systems provide limited benefits in this scenario, struggling to effectively integrate both conversational and repository context.

On the methodological side, a limited number of existing studies have explored this problem from different perspectives.
For example, \citet{liu2026scalable} extend memory mechanisms centered solely on conversation, building upon the general memory system Mem0 \cite{chhikara2025mem0} and proposing Mem0$^\mathcal{R}$ to integrate conversational history with relevant code context in repository settings; 
CodeMEM \cite{wang2026codemem} emphasizes structure-based modeling on the code side. It incorporates abstract syntax trees to guide memory construction and updating, thereby enhancing the representation of code semantics.

Overall, research on memory systems for repository-oriented conversation is still at an early stage. Existing approaches typically focus on a single perspective,
such as conversational information, code structure, or repository history, and lack a unified representation and coordinated utilization of multiple information sources. 
This highlights the need to integrate contextual information from different sources and at different granularities, which is exactly what \ourmethod{} aims to address.

\section{Conclusion}

In this work, we investigated how conversational history and repository context can be effectively managed together in repository-oriented multi-turn interactions. \ourmethod{} addresses this problem by maintaining their associations through a dual memory graph and selectively recalling complementary conversational and code context for each request. Evaluations on LoCoEval and CodeAssistBench show that this design improves code generation and problem-solving effectiveness across both controlled long-horizon conversations and realistic project-grounded interactions. The retrieval analysis further shows that natural-language and code-side retrieval provide asymmetric but complementary evidence, while the ablation results confirm that graph-based retrieval and direct use of repository structure make substantial contributions to generation accuracy.

More broadly, our findings suggest that repository-oriented context management should not be viewed merely as a mechanism for compressing histories that exceed an LLM's context window. Even when the complete conversation remains manageable, selectively organizing and retrieving relevant context can improve interaction effectiveness. As conversations grow longer, the same mechanism additionally limits the increasing cost of repeatedly supplying accumulated history. Repository-oriented context management is therefore fundamentally a problem of maintaining the relationship between evolving conversational intent and repository state, and exposing the information relevant to the current task. \ourmethod{} provides one realization of this perspective by explicitly preserving and jointly utilizing these two sources of context.

\bibliographystyle{ACM-Reference-Format}
\bibliography{refs}

\end{document}